# NEW APPROACH TO THE ESTIMATION OF THE LOWEST BOUNDARY OF AN AXION MASS AND AXION-ELECTRON COUPLING CONSTANT

**Sevinj O. Guseynova[1†], Vali A. Huseynov[2,1,3,4*††], Mehriban Sh. Hasanova[5], Sevinj Y. Rzayeva[6], Nigar E. Musazade[7]**

[1] *Laboratory for Cosmic Rays, Nuclear and High Energy Physics, Institute of Physics, H. Javid Ave., 131, Baku, AZ 1073, Azerbaijan;*
[2]*Department of Physics, Sumgayit State University, Baku Str., 1, Sumgayit, AZ 5008, Azerbaijan;* vgusseinov@yahoo.com;
[3]*Department of Physics, Baku Engineering University, H. Aliyev Str., 120, Khirdalan, AZ 0101, Azerbaijan;* vehuseynov@beu.edu.az;
[4] *Scientific Research Center, Odlar Yurdu University, Koroghlu Rahimov Str., 13, Baku, AZ 1072, Azerbaijan;*
[5] *Department of Engineering Physics and Electronics, Azerbaijan Technical University, H. Javid Ave., 25, Baku, AZ 1073, Azerbaijan;*

[6] *Department of Electroenergetics Engineering, Nakhchivan State University, University Campus, Nakhchivan, AZ 7012, Azerbaijan*

[7] *Department of General Physics, Azerbaijan State Pedagogical University, U. Hajibayli Str., 68, Baku, AZ 1000, Azerbaijan;*

† The author's name in her other papers is also S. O. Huseynova or S.O. Tagieva.

†† The author's name in his other papers is also V. A. Gusseinov or V. A. Guseinov.

*Corresponding author: vgusseinov@yahoo.com; vehuseynov@beu.edu.az

**Abstract** In this work, we have performed the theoretical estimation of the lowest boundary of an axion mass. We have analyzed the energy spectrum of the electron in a constant homogeneous magnetic field, taking into account its anomalous magnetic moment. We have applied the obtained results on the energy spectrum of the electron in a constant homogeneous magnetic field with allowance for its anomalous magnetic moment to the hydrogen atom electron, which is located in addition to the Coulomb field of the proton, as well as in the magnetic field produced by the magnetic moment of the proton. We have applied the idea of the restriction for the lowest boundary of the axion mass to the triplet-singlet transition between two adjacent levels obtained as a result of the hyperfine structure splitting of the $s$-levels in the hydrogen atom. We have concluded that the mass of an axion ought to be more than $5.885 \times 10^{-6}\ eV$. We have also obtained the analytical relation for the lowest boundary of the axion-electron coupling constant and numerically estimated its value: the axion-electron coupling constant ought to be more than $1.670 \times 10^{-16}$.

## 1. Introduction

An axion being one of the cold dark matter candidates was theorized by Weinberg [1] and Wilczek [2]. As the pseudo-Goldstone boson, an axion solves the problem connected with the strong breaking of $CP$ invariance in the Standard Model. An axion appears in the results of the spontaneous breaking of the global $U(1)_{PQ}$ Peccei-Quinn symmetry [3]. In the standard axion model [1], the energy scale $f_a$ of violation of the Peccei-Quinn symmetry coincides with the electroweak scale $v_w = \left(\sqrt{2}G_F\right)^{-1/2} \cong 250\ GeV$ where $G_F$ is the Fermi constant. In contrast to the classical theory, in quantum theory, an axion acquires a tiny mass due to its interaction with gluons [1]. The above-indicated standard axion model was excluded by the experiment [4]. Subsequently, various models of the "invisible" axion in which $f_a \gg v_w$ were developed [5, 6].

One of the most important problems existing in Particle Physics, Astrophysics and, especially, in Cosmology is associated with the determination of the axion mass. The axion mass is given by the relation $m_a = [(5.70 \pm 0.07) \times 10^6\ eV]GeV/f_a$ [7, 8] where $f_a$ is the QCD parameters. This relation depends on specific axion models [8-13]. The plausible range of the masses of an axion which is characteristic for post-inflationary scenarios is $m_a \approx 2.6 \times 10^{-5} eV - 5 \times 10^{-4} eV$ if one considers the axion models with short-lived domain walls [14-16], such as the Kim-Shifman-Vainshtein-Zakharov (KSVZ) model [17, 18] with $N_{DW} = 2N_Q = 1$. In case of the Dine-Fischler-Srednicki-Zhitnitsky (DFSZ) model [19, 20] with long-lived domain walls ($N_{DW} > 1$), the predicted mass is $m_a \gtrsim 10^{-3} eV$ [21, 22]. The axion field oscillations about the effective potential minimum generate a cosmological population consisting of cold axions whose abundance depends on the axion mass [23-25]. If the axion mass is $m_a \gtrsim 5 \times 10^{-6} eV$ , it could be both a dark-matter candidate and solve the strong CP problem. If inflation took place at a low scale and lasted sufficiently long, the mass of an axion could be $m_a \approx 1\, peV$ [13, 26, 27]. Simulating the formation of axions during the post-inflation period, Borsanyi et al [28] reported that an axion mass is between 0.05 and 1.50 $meV$. Estimations used numerical simulations on the present abundance of a KSVZ-type axion lead to values between 0.02 and $0.1\, meV$ [29, 30]. The Axion Dark Matter Experiment excluded the existence of the axion with the masses in the ranges $1.9 - 3.53\ \mu eV$ [31] and $3.3 - 4.2\ \mu eV$ (for the KSVZ model) [32, 33]. Based on the analysis of four neutron stars, an upper limit on the axion mass of $0.079\ eV$ was obtained by Berenji et al [34]. The results obtained from the experiment based on the method proposed by a theoretical team from the Massachusetts Institute of Technology exclude the existence of axions in the mass range from $4.1 \times 10^{-10}\ eV$ to $8.27 \times 10^{-9}\ eV$ [35]. The results of the polarized light measurements performed by the Event Horizon Telescope rejected the approximate $10^{-21} eV - 10^{-20}\ eV$ range for the axion mass values [36]. The Magnetized Disk and Mirror Axion experiment is based on the concept of a dielectric haloscope for the direct search of dark matter axions in the mass range of 40 to 400 $\mu eV$ [37]. The results on an axion dark matter search at DFSZ sensitivity with the CAPP-12TB haloscope excluded specific axion-photon couplings in the 4.51 $\mu eV$ to 4.59 $\mu eV$ range [38]. An ultralight axion with $m_a \sim 10^{-22}\ eV$ seems to solve the small-scale problems of cold dark matter. The correct relic density without fine-tuning is provided by a single ultralight axion with a GUT scale decay constant [39]. Using the upper and lower bounds on $f_a$ from cosmology and astrophysics, the QCD axion mass is allowed in the range between $6 \times 10^{-6}\ eV$ and $2 \times 10^{-3}\ eV$ [40]. In [41], the authors computed higher order corrections to the axion mass and studied the dependence of the axion mass on the temperature, which affects the axion relic abundance today. The review article [8] is devoted to the current status of constraints on axions from stellar physics with the focus in particular on the Sun, globular cluster stars, white dwarfs and (proto)-neutron stars.

The additional data on axion, axion mass and other related questions can be found from the review articles on axions (see: e.g., [8, 42, 43]).

Nowadays, a wide variety of complementary technologies developed by theoretical and experimental groups worldwide is capable to detect QCD axion dark matter across a significantly broader mass range: $10^{-11}\ eV \lesssim m_a \lesssim 10^{-2}\ eV$ [44]. However, despite great successes in the exploration of the axion mass, the problem associated with the lowest and uppermost boundaries of the axion mass, has not been resolved yet and this problem remains as an important puzzle of modern Particle Physics, Astrophysics and Cosmology.

In the presented work, we are interested in the following questions: (i) how can theoretically be determined the lowest boundary of the axion mass? (ii) how much is the value of the lowest boundary of the axion mass? Namely, the answers to these questions are the main motivations of the presented research.

The presence of the term

$$\mathcal{L}_{af} = C_{af}\frac{\partial_\mu a}{2f_a}\left(\bar{\psi}_f\gamma^\mu\gamma^5\psi_f\right) = \frac{g_{af}}{2m_f}\partial_\mu a\left(\bar{\psi}_f\gamma^\mu\gamma^5\psi_f\right) \tag{1}$$

in the axion interaction Lagrangian with the Standard Model fields [8] indicates that an axion can couple with fermions where $g_{af} = C_{af}m_f/f_a$ is the dimensionless coupling constant, $C_{af}$ is a numerical coefficient depending on the specific model, $f_a$ is a model dependent energy scale (which is called the PQ (or axion) constant), $m_f$ is the fermion mass, $\gamma^\mu$ and $\gamma^5 = -\gamma^0\gamma^1\gamma^2\gamma^3$ is the Dirac matrices, $\psi_f$ is the fermion wave function, $\bar{\psi}_f = \psi_f^+\gamma^0$. In fact, the DFSZ-type axions have direct, tree-level couplings to leptons while the KSVZ-type axions do not interact with leptons at the tree level [8]. To the first order in $a/f_a$, the Lagrangian given by the expression (1) can be replaced with the pseudoscalar interaction Lagrangian (see: e.g., [5, 6, 45])

$$\mathcal{L}_{af} = -ig_{af}\left(\bar{\psi}_f \gamma^5 \psi_f\right)a. \tag{2}$$

If we take an electron as a fermion in the Lagrangian (2), we get

$$\mathcal{L}_{ae} = -ig_{ae}\left(\bar{\psi}_f \gamma^5 \psi_f\right)a \tag{3}$$

where $g_{ae}$ is the constant responsible for the axion-electron coupling [8]) which induces several axion production processes (see: e.g., [5, 6, 8]). One of the axion production processes is the emission of an axion ($a$) by an electron in a magnetic field (the axion synchrotron emission)

$$e^- \to e^{-\prime} + a \tag{4}$$

which is quite possible according to the Lagrangian (3). The axion synchrotron emission in a magnetic field was studied in the papers [45, 46]. In [45], the authors calculated the contribution of the axion synchrotron emission by an electron to the axion luminosity of a magnetized, highly degenerate relativistic electron gas and obtained a new bound on the axion electron coupling constant: $g_{ae} \lesssim 5 \times 10^{-14}$. In [46], the authors considered the process (4) proceeding by virtue of an intermediate plasmon even if axions do not couple to electrons at the tree level and calculated the axion emissivity of a plasma in the presence of a strong magnetic field due to the resonant transition between longitudinal plasmons and axions.

The main purpose of the presented work is theoretically to determine the lowest boundary of the axion mass. To achieve this purpose, we consider the axion synchrotron emission by an electron (4) at the tree level in a constant and homogeneous magnetic field with allowance for the anomalous magnetic moment of an electron. The main idea of determining the axion mass is based on the comparison of the energy differences in three physical cases (or phenomena) which are densely connected each other: (i) the energy difference between two adjacent Landau sublevels of an electron with allowance for its anomalous magnetic moment and spin; (ii) the energy difference in the reversal of the electron magnetic moment in an external magnetic field; (iii) the energy difference corresponding to the triplet-singlet transition between the energy levels arisen in the result of the hyperfine structure splitting of the $s$-levels in the hydrogen atom. When the energy difference between the the initial state and final state electrons participating in the reaction (4) is not less than the axion mass,

$$\Delta E = E - E' \geq m_a \tag{5}$$

then an axion emission by the electron is quite realistic where $m_a$ is the axion mass and $E$ ($E'$) is the energy of the electron in the initial (final) state. However, the condition

$$m_a > (\Delta E)_{max} \tag{6}$$

puts a limit on the emission of an axion by the electron situated in an external magnetic field. The inequality (6) shows that the lowest boundary of the axion mass is determined by the maximum value of $\Delta E$. This means that we are namely interested in searching for such two adjacent energy levels or sublevels for which the energy difference $\Delta E$ is maximum.

**2. Energy difference between two adjacent Landau sublevels of an electron**

We start to analyze the energy spectrum of the electron in a constant homogeneous magnetic field with allowance for its anomalous magnetic moment and spin. The energy spectrum of an electron in a constant homogeneous magnetic field with allowance for its anomalous magnetic moment, spin and third component of momentum is given by [47-48]

$$E_n(\zeta,\, p_z,\, H) = m_e \left\{ \left(\frac{p_z}{m_e}\right)^2 + \left( \sqrt{1 + \frac{H}{H_0}(2n_L + \zeta + 1)} + \zeta \frac{H}{H_0} \frac{\mu_e - \mu_0}{2\mu_0} \right)^2 \right\}^{1/2} \tag{7}$$

where $p_z$ is the electron momentum $z$-component along the magnetic field direction, $n_L = 0, 1, 2, \ldots$ is the number of the Landau energy level, $\zeta = \pm 1$ characterizes the election spin projection onto the magnetic field direction,

$$\mu_e = \mu_0(1 + a_e), \quad (8)$$

is the electron magnetic moment containing both the normal (Dirac) magnetic moment $\mu_0 = e/(2m_e) \simeq 5.79 \times 10^{-9}\ eV/G$ (the Bohr magneton) and the anomalous magnetic moment $\Delta\mu_e = \mu_0 a_e$,

$$a_e = \frac{g-2}{2} = \frac{\alpha}{2\pi} + \cdots, \quad (9)$$

$g$ is the Lande factor [30-33], $H_0 = m_e^2/e = 4.41 \times 10^{13} Oe = 4.41 \times 10^{13}\ G$ is the Schwinger field strength of an external magnetic field, $m_e$ is the electron mass, $e$ is the elementary electric charge, $e > 0$, $\alpha = e^2 \approx 1/137.04$ is the fine-structure constant. In this work, we use the system of units $\hbar = c = 1$.

If we assume that the $p_z$ component of the electron momentum is equal to zero or is negligible small, then we obtain the following expression for the energies of the $n$-th Landau level:

$$E_n(\zeta, f) = m_e \left[\sqrt{1 + f(2n_L + \zeta + 1)} + \frac{1}{2} a_e f \zeta\right], \quad (10)$$

where $f = H/H_0$ is the dimensionless field parameter. Depending on the electron spin polarization, each of the above indicated Landau levels, except the ground Landau level, has two sublevels corresponding to the spin polarizations $\zeta = +1$ and $\zeta = -1$. The energies corresponding to the sublevels of $(n_L, \zeta = +1)$ and $(n_L, \zeta = -1)$ of the electron with allowance for its anomalous magnetic moment are presented in Table 1.

Table 1. Energies of the adjacent Landau sublevels of the electron with allowance for its anomalous magnetic moment

| Number of the Landau levels | Sublevel | Energy of the sublevel |
|---|---|---|
| $n_L$ | $n_L, \zeta = +1$ | $E = E_{n_L}^{+1} = m_e\left[\sqrt{1+2f(n_L+1)} + \frac{a_e}{2}f\right]$ |
| | $n_L, \zeta = -1$ | $E' = E_{n_L}^{-1} = m_e\left[\sqrt{1+2fn_L} - \frac{a_e}{2}f\right]$ |

It is visible from Table 1 that the energy of the Landau sublevel corresponding to the quantum state $|n_L, \zeta = +1\rangle$ is more than the energy of the Landau sublevel corresponding to the quantum state $|n_L, \zeta = -1\rangle$

$$E_{n_L}^{+1} > E_{n_L}^{-1} \tag{11}$$

This inequality is true for the arbitrary Landau level satisfying the condition $n_L \geq 1$. The electron occupying the ground Landau level is in the non-degenerate state: $n_L = 0$, $\zeta = -1$. Therefore, the inequality (11) is not applicable for the electron occupying the ground Landau level.

Using the Table 1, we obtain for energy difference between two adjacent sublevels $E_{n_L}^{+1}$ and $E_{n_L}^{-1}$:

$$\Delta E_L = E - E' = E_{n_L}^{+1} - E_{n_L}^{-1} = (1 + a_e)fm_e. \tag{12}$$

The difference $\Delta E_L$ given by the formula (12) is the maximum value of $\Delta E_L$, i.e.

$$(\Delta E_L)_{max} = (1 + a_e) f m_e = (1 + a_e)\frac{e}{m_e}H. \qquad (13)$$

## 3. Reversal of the electron magnetic moment in an external magnetic field

Potential energy of interaction of the electron magnetic moment $\vec{\mu}_e$ with the external magnetic field $\vec{H}$ is given by

$$U = -\vec{\mu}_e\vec{H}. \qquad (14)$$

The related torque in an external magnetic field

$$\vec{\tau} = \vec{\mu}_e \times \vec{H} \qquad (15)$$

makes the electron magnetic moment to rotate from the state (when the angle $\theta$ between the electron magnetic moment and magnetic field vector $\vec{H}$ is $\theta = \pi$) characterizing with the highest potential energy

$$U_2 = -\mu_e H \cos 180^{\circ} = \mu_e H \qquad (16)$$

to the state ($\theta = 0$) characterizing with the lowest potential energy

$$U_1 = -\mu_e H \cos 0 = -\mu_e H. \qquad (17)$$

In this reversal of the electron magnetic moment in an external magnetic field, the potential energy difference $\Delta U = U_2 - U_1$ is maximum

$$(\Delta U)_{max} = 2\mu_e H = (1 + a_e)\frac{e}{m_e}H. \qquad (18)$$

So, the result (13) obtained in Section 2 and the result (18) obtained in this Section coincide and they confirm that the maximal energy emitted in the transition between two adjacent Landau sublevels is

$$(\Delta E_L)_{max} = (\Delta U)_{max} = m_a^{lb} \qquad (19)$$

where $m_a^{lb}$ denotes the lowest boundary of the axion mass. Using the equality (19), we can also write the relation (6) as follows

$$m_a > m_a^{lb} \quad (20)$$

where

$$m_a^{lb} = (1 + a_e)\frac{e}{m_e}H. \quad (21)$$

The relation (21) means that the restriction for the axion mass is determined by the possible maximum value of the magnetic field strength $H$. At this stage, we can not directly determine the possible maximum value of the magnetic field strength $H$. We will determine it a bit later.

To determine the lowest boundary of the axion mass, we consider the hyperfine structure splitting phenomenon in the hydrogen atom.

## 4. The electron in the magnetic field produced by the proton magnetic moment in a hydrogen atom and hyperfine structure splitting

A typical example for the electron in a magnetic field is the hydrogen atom electron which is located, in addition to the Coulomb field of the proton, as well as in the magnetic field produced by the proton magnetic moment

$$\mu_p = k\mu_N \quad (22)$$

where $k \cong 2.793$ and

$$\mu_N = \frac{e}{2m_p} \cong 3.152 \times 10^{-12}\frac{eV}{G} \quad (23)$$

is the nuclear magneton, $m_p \cong 938.272\ MeV$ is the proton mass [49]. The proton magnetic moment $\vec{\mu}_p = \mu_p\vec{\sigma}_p'$ produces the vector potential [50, 51]

$$\vec{A} = curl\ \frac{\mu_p \vec{\sigma}'_p}{r} \tag{24}$$

that determines the magnetic field vector

$$\vec{H} = curl\ \vec{A} = \mu_p \frac{3\vec{r}(\vec{\sigma}'_p \vec{r}) - \vec{\sigma}'_p r^2}{r^5} \tag{25}$$

where $\vec{\sigma}'_p$ are the proton spin matrices [50]. It is known that the hyperfine structure of the spectral lines in the hydrogen atom is connected with the interaction of the proton magnetic moment $\vec{\mu}_p$ with the electron magnetic moment $\vec{\mu}_e = -\mu_0 \vec{\sigma}'_e$ where $\vec{\sigma}'_e$ is the electron spin matrices. In fact, the magnetic field produced by the proton magnetic moment (the formula (25)) affects on the electron magnetic moment, resulting in an additional interaction between the proton and electron of the hydrogen atom that leads to the hyperfine structure splitting of the $s$-levels in the hydrogen atom [50]

$$\Delta E_S = \frac{8}{3} \mu_e \mu_p \frac{1}{n^3 a_0^3} [2S(S+1) - 3] \tag{26}$$

where $n$ is the principal quantum number ($n = 1,2,3, \ldots.$) of the hydrogen atom electron,

$S$ is the total spin of the proton and electron in the hydrogen atom, $a_0 = 1/(m_e e^2)$ is the first Bohr orbit radius.

The main idea for the determination of the lowest boundary of the axion mass is based on the triplet-singlet transition between the energy levels arisen in the result of the hyperfine structure splitting. It is obvious that the axion mass can not be equal to or less than the maximum difference between the two energy levels formed after hyperfine structure splitting. If the axion mass were

$$m_a \leq \left(\Delta E_{hf}\right)_{max}, \tag{27}$$

then the hydrogen atom electron would continuously emit an axion at the expense of the interaction between the electron magnetic moment and the magnetic field produced by the proton magnetic moment in the the hydrogen atom which is impossible because of the stability of the hydrogen atom where

$$\Delta E_{hf} = \Delta E_S(S=1) - \Delta E_S(S=0). \tag{28}$$

Therefore, the axion mass must satisfy the condition

$$m_a > \left(\Delta E_{hf}\right)_{max}. \tag{29}$$

## 5. Constraint for the lowest boundary of the axion mass

The energy difference $\Delta E_{hf}$ corresponding to the triplet-singlet transition between the hydrogen energy levels $\Delta E_S(S=1)$ and $\Delta E_S(S=0)$ arisen in the result of the hyperfine structure splitting of the $s$-levels is given by the formula (see: [50])

$$\Delta E_{hf} = \frac{8}{3}\alpha^4 k_p (1+a_e)\frac{m_e^2}{m_p}\frac{1}{n^3}\,. \tag{30}$$

In this formula the anomalous magnetic moment of an electron has been taken into account. When the principal quantum number is $n = 1$, the quantitity $\Delta E_{hf}$ is maximum

$$\left(\Delta E_{hf}\right)_{max} = \frac{8}{3}\alpha^4 k_p (1+a_e)\frac{m_e^2}{m_p}\,. \tag{31}$$

The energy differences $(\Delta E_L)_{max}$, $(\Delta U)_{max}$ and $\left(\Delta E_{hf}\right)_{max}$ arise from the transition based on the same physical phenomenon. Therefore, the energy differences $(\Delta E_L)_{max}$, $(\Delta U)_{max}$ and $\left(\Delta E_{hf}\right)_{max}$ are equal each other

$$(\Delta E_L)_{max} = (\Delta U)_{max} = \left(\Delta E_{hf}\right)_{max} = m_a^{lb} \tag{32}$$

Using the relations (20), (31) and (32), we obtain

$$m_a > \frac{8}{3}\alpha^4 k_p (1+a_e)\frac{m_e^2}{m_p} \tag{33}$$

If the energy difference $\left(\Delta E_{hf}\right)_{max}$ given by the relation (31) satisfies the condition (20) or (29), it means that gaining an additional energy at the expense of the interaction between the electron magnetic moment and the magnetic field produced by the proton magnetic moment in the the hydrogen atom, the electron can not emit an axion of the mass heavier than

$$m_a^{lb} = \left(\Delta E_{hf}\right)_{max} = \frac{8}{3}\alpha^4 k_p (1 + a_e)\frac{m_e^2}{m_p} \tag{34}$$

where $m_a^{lb}$ is the lowest boundary of the axion mass. Otherwise, if the axion mass were not more than the energy difference $\left(\Delta E_{hf}\right)_{max}$ given by the formula (31), i.e., if it were $m_a \leq m_a^{lb}$, then the hydrogen atom electron would continuously emit an axion at the expense of the interaction between the electron magnetic moment and the magnetic field produced by the proton magnetic moment in the the hydrogen atom which is impossible because of the stability of the hydrogen atom. So, the numerical estimation obtained from the relations (33) and (34) yields the lowest boundary for the axion mass: $m_a > m_a^{lb} \cong 5.885 \times 10^{-6}\ eV$. It should be noted that the lowest boundary for the axion mass is expressed through the fundamental constants (see: the formula (34)).

## 6. Axion-electron coupling constant

If we take into account the relations (33) and (34) which we obtained for the lowest boundary of an axion mass and the following known relation [5, 6]

$$g_{ae} = 2.83 \times 10^{-11}\frac{m_a \cos^2\beta}{eV}, \tag{35}$$

we obtain the following relation for the lowest boundary of the axion-electron coupling constant

$$g_{ae} > g_{ae}^{lb} = 7.55 \times 10^{-11}\alpha^4 k_p (1 + a_e)\frac{m_e^2}{m_p}\frac{m_a \cos^2\beta}{eV}. \tag{36}$$

According to the DFSZ model $\cos\beta \approx 1$. If we take into account $\cos\beta \approx 1$ in the relation (36), we obtain the following numerical estimation for the lowest boundary of the axion-electron coupling constant

$$g_{ae}^{lb} \cong 1.67 \times 10^{-16} . \qquad (37)$$

This means that

$$g_{ae} > g_{ae}^{lb}. \qquad (38)$$

**7. Conclusion**

We have started from the idea that an electron located in an external magnetic field can emit an axion. The emission of an axion by an electron in an external field is not allowed by the energy conservation law if the energy difference between two Landau sublevels is less than the axion mass. This puts the restriction for the lowest boundary of the axion mass. We applied the idea of the restriction for the lowest boundary of the axion mass to the triplet-singlet transition between two adjacent levels obtained in the result of the hyperfine structure splitting of the $s$-levels in the hydrogen atom. We have also taken into account the electron anomalous magnetic moment in our calculations. The maximum value of the energy difference between the level with $S = 1$ and the level with $S = 0$ is provided when $n = 1$. This enables us to establish the relation for determination of the lowest boundary of the axion mass. Despite, in general, the axion synchrotron emission by the electron located in an external magnetic field is allowed, the strength of the magnetic field produced by the magnetic moment of the proton in the hydrogen atom is not sufficient for emission of an axion by the hydrogen electron. It enables us to conclude that the mass of an axion ought to be more than $5.885 \times 10^{-6}\ eV$. Otherwise, if the axion mass were less than the estimated lowest boundary ($5.885 \times 10^{-6}\ eV$), then the hydrogen atom electron would continuously emit an axion which contradicts to the stability of the hydrogen atom. We have also obtained the analytical relation for the lowest boundary of the axion-electron coupling constant and numerically estimated its value: the axion-electron coupling constant ought to be more than $1.670 \times 10^{-16}$.